# Solving non-linear equations of longitudinal and transverse electron waves in collisionless Maxwellian plasma[1]

V. N. Soshnikov[2]

Plasma Physics Dept.,
All-Russian Institute of Scientific and Technical Information
of the Russian Academy of Sciences
*(VINITI, Usievitcha 20, 125315 Moscow, Russia)*

**Abstract**

We have considered an expansion of solutions of the non-linear equations for both longitudinal and transverse waves in collisionless Maxwellian plasma in series of non-damping overtones of the field $E(x,t)$ and electron velocity distribution function $f = f_0 + f_1$ where $f_0$ is background Maxwellian distribution function; $f_1(\vec{v}, v_x, v_z, x, t) = \sum_{n=1} F_n(\vec{v}, v_x, v_z) \cos[n(\omega t - kx + \varphi_1)]$; the electrical field is $E(x,t) = \sum_{n=1} E_n(x,t) \cos[n(\omega t - kx + \varphi_1) \pm \pi/2]$ (depending on the sign of $\mp |E_1|$); amplitudes $E_n$, $F_n$ are proportional to $(E_1)^n$, what provides the convergence of expansion at any small enough $|E_1|$. It is shown presence of recurrent relations for arising overtones. There are proposed also procedures of cutting off the distribution function $|f_1|$ to the condition values $0 \le f \le 2 f_0$ near the singularity points where kinetic equation becomes inapplicable. In this case, at poles absence the solution reduces to non-damping Vlasov waves (oscillations). In the case of transverse waves, dispersion equation has two roots, corresponding to the branches of fast electromagnetic and slow electron waves. There is noted a possibility of experimental testing appearing exotic results of overtones availabilities.

PACS numbers: *52.25 Dg; 52.35 Fp.*
Key words: plasma oscillations; plasma waves; Landau damping; recurrent relations for overtones.

### 1. Introduction

Landau theory of waves in collisionless Maxwellian plasma [2] is result of the only Landau work in this field and commences the whole physical trend named Landau damping. This theory went down in all plasma physics courses, text-books, reviews, into their plasma wave sections as an indisputable fact (see for example [3 – 8]). Landau damping as a firmly settled conception is cited and used in thousands papers with its diverse developments and applications. Ones consider it as proved experimentally and unquestioned in all cases.

The essence of Landau theory [2], [3] reduces to calculation logarithmically divergent integrals in linearized dispersion equations of longitudinal electrostatic waves (Vlasov, [9]) or transverse electromagnetic and electron waves not in the principal value sense with finite dispersion integrals [9] but in the sense of contour integrals in velocity $v_x$ with additional bypassing around the poles $v_x = \pm \omega/k$ on the real value axis $v_x$ after analytical continuation of $v_x$ into complex value plane [3] and complex extension of real wavenumber $k$ or frequency $\omega$ with using complex wave function. Thus, in Landau theory, to the dispersion now complex integral in the principal value sense (Vlasov) one adds imaginary residuum at a singularity point $v_x = \pm \omega/k$ with its bypassing around along the half-circle in complex plane $v_x$.

---

However already in the some later works ones noted difficultly explicable paradoxes of Landau theory (see, par example, [4] and further references [10] –[14]): unavoidable presence together with exponentially damping waves also exponentially growing ones (cf. [10]); the presence of non-damping kinematical waves at simultaneously exponentially damping electrical field [4], [10]; possibility of negative total distribution function $f_0(\vec{v}) + f_1(\vec{v}, v_x)$ with Maxwellian function $f_0$ due to singularity point $v_x = \pm \omega/k$; the indefiniteness at calculation of residuum in the real-value pole $v_x = \pm \omega/k$ (that is either one ought to take half-circle contour at bypassing around the pole [2], [3], or the whole circle [11], since then at solving dispersion equation (9a) or its equivalent (9b), the real value poles $\pm \omega/k$ collisionally remove at $\text{Im}\, k \neq 0$ arbitrarily small into the complex plane). Thus the poles $v_x = \pm \omega/k$ with an infinitely small imaginary collisional addition in the wave solution with the decrement/increment term $\pm \text{Im}\, k$ (in the case of boundary problem) shift upward or downward correspondingly to the direction of the poles bypassing around clockwise or counter clockwise. But the simultaneous presence of damping/growing waves without dissipative energy transfer at collisions between electrons $f_1$ and $f_0$ violates the law of energy conservation. Besides it, distribution function $f = f_0 + f_1$ can be infinite and negative (see also [14].

Sometimes used qualitative grounds of Landau damping based on figurative considerations of monoenergetic electron beams with velocities more or less than wave phase velocity $\pm \omega/k$ appear unconvincing due to the real observable value is wave group velocity which is, for example, for transverse electromagnetic wave less than light velocity $c$, whereas phase velocity is more than light velocity and appears only a convenient mathematical abstraction.

Appearing of indefinitely divergent dispersion integrals in wave solution is a result of the information lack in original kinetic and field equations neglecting collision terms, thus additional physical conditions are required which have been formulated and can be better satisfied with the indefinitely divergent integrals taken in the principal value sense [14] without addition Landau poles.

The radical decision relative to versions both in the sense of Vlasov rule (dispersion integral in the principal value sense), and Landau rule (complex extended dispersion integral in the principal value sense plus residua at the poles $v_x = \pm \omega/k$) is a physically reasonable cutting off $f_1(v_x) \to \pm \infty$ near the point of singularity with setting values to be $0 < f_0 + f_1 < 2 f_0$. Function $f_1$ can not be larger $f_0$ because then in opposite wave phases $f = f_0 + f_1$ can be periodically negative.

By this inequality at the same time it is achieved the physical justification of finiteness of the distribution function and dispersion integrals without the need to use directly both versions and complex additives of collisionless Landau damping. Thus it is concluded [14] that Landau damping as a real physical process does not exist.

This is due to using at derivation the dispersion relation the complex wave functions, respectively, with searching its solution in the form of a complex root of the complex dispersion equation with two parameters: the wave number $\text{Re}\, k = k_1$ and the simultaneously both decrement/increment $\text{Im}\, k = \pm k_2$, whereas physically justified is finding solutions in the form of real waves and real electron distribution function. In this case we get a real dispersion equation averaged over oscillation period

$$1 = -\Phi(k_1, k_2) \cdot \omega_0^2 \int \frac{\partial f_0 / \partial v_x}{(\omega - k_1 v_x)} dv_x; \quad \Phi \equiv \int_0^{2\pi} \frac{dy}{k_1 + k_2 \tan y}$$

(where integrand in rather specific integral $\Phi$ contains singular points and can not be taken analytically, with the need for special consideration and numerical calculation), to find that it is necessary to add the second equation of conservation energy law with parameter $k_1$ and damping parameter $k_2$ [14].

## 2. Preliminary consideration in linear approach

The following further consideration in this work is limited with some principal features of the simplest equations of wave propagation in half-infinite homogeneous collisionless Maxwellian plasma slab, correspondingly for longitudinal waves (self-consistent Vlasov equations):

$$\frac{\partial f_1}{\partial t} + v_x \frac{\partial f_1}{\partial x} - \frac{|e| E_x(x,t)}{m_e} \frac{\partial f_0}{\partial v_x} = \frac{|e| E_x(x,t)}{m_e} \frac{\partial f_1}{\partial v_x} \sim 0, \tag{1}$$

$$\frac{\partial E_x(x,t)}{\partial x} = -4\pi |e| n_e \int f_1 d\vec{v}, \tag{2}$$

and transverse ones

$$\frac{\partial f_1}{\partial t} + v_x \frac{\partial f_1}{\partial x} - \frac{|e| E_z(x,t)}{m_e} \frac{\partial f_0}{\partial v_z} = \frac{|e| E_z(x,t)}{m_e} \frac{\partial f_1}{\partial v_z} \sim 0, \tag{1a}$$

$$\frac{\partial^2 E_z(x,t)}{\partial x^2} - \frac{1}{c^2} \frac{\partial^2 E_z(x,t)}{\partial t^2} + \frac{4\pi |e| n_e}{c^2} \frac{\partial}{\partial t} \int v_z f_1 d\vec{v} = 0, \tag{3}$$

where

$$f_0 = \left(\frac{m_e}{2\pi k_B T}\right)^{3/2} e^{-\frac{mv^2}{2k_B T}}, \tag{4}$$

$k_B$ is Boltzmann constant, $m_e$ is electron mass; further $e \equiv |e|$ is positive value of negative electron charge; $E_x$ or $E_z$ are replaced further for simplicity by $E$; $x$ is coordinate along wave propagation. The electron velocity distribution function $f$ is presented as a sum of Maxwellian part $f_0(v)$ normalized to unity and small perturbation $|f_1(v,v_x,v_z,x,t)| \ll f_0$; linearization is reduced to removing small quadratic in perturbations right hand side terms (r.h.s.) of equations (1) and (1a).

Discussion on solution properties of the linearized wave equations (1), (2) with integrals in the consequent dispersion equation taken in the principal value sense [12] results in elementary real value non-damping asymptotical solution at the boundary condition $E(x=0, t) = E_0 \cos \omega t$ of the type

$$E(x,t) = E_0' \cos(\omega t - kx + \psi), \tag{5}$$

$$f_1(v,v_x,x,t) = E_0' F(v,v_x) \cos(\omega t - kx + \varphi), \tag{6}$$

$$\varphi = \psi - \pi/2, \tag{7}$$

$$F(v,v_x) = \frac{e}{m_e} \frac{\partial f_0 / \partial v_x}{\omega - kv_x}, \tag{8}$$

with the traditional dispersion equation with real $k$

$$1 = -\omega_L^2 \frac{1}{k} \int \frac{\partial f_0 / \partial v_x}{\omega - kv_x} dv_x \tag{9}$$

where $\omega_L = \left(4\pi e^2 n_e / m_e\right)^{1/2}$ is Langmuir frequency, with real-value roots $k(\omega)$ if it is taken in the principal value sense. The integration limits are determined by given boundary conditions.

The longitudinal waves excited by electric field $E_x(x=0) \sim \cos \omega t$ perpendicular to plane $x = \pm 0$ in half-infinite $0 < |x| < \infty$ collisionless Maxwellian plasma are non-damping. The proposed solutions can be oppositely directed from the point $x = 0$ waves at $E_x' = $ const :

$$\begin{cases} E_x(x,t) = E_x' \cos(\omega t - kx + \varphi), & F = F(\omega - kv_x + \psi) \quad \text{at } x > 0 \\ E_x(x,t) = E_x' \cos(\omega t + kx + \varphi), & F = F(\omega + kv_x + \psi) \quad \text{at } x < 0 \end{cases}$$

without violation of energy conservation for non-damping longitudinal electron wave solutions with real $k$.

In the case of a plasma with reflecting walls at large $|x|$, for example with mirror symmetric in $x = 0$ (and relatively to $v_x = \pm \omega/k$) solutions of the type (6), at calculating the integrals of the type (9) with the account of the integral in principle value sense, dispersion equation has no singularities, including the point $v_x$ near to $\omega \pm kv_x \sim 0$. If the solution $k(\omega)$ is real value, then complete solution in this case is usual standing waves (oscillations):

$$\begin{cases} f_1(x,t) \propto \left[\cos(\omega t - k_1 x + \varphi_1) + \cos(\omega t + k_1 x + \varphi_2)\right] = 2\cos(\omega t + \frac{\varphi_1 + \varphi_2}{2}) \cdot \cos(k_1 x + \frac{\varphi_1 - \varphi_2}{2}), \\ E(x,t) \propto 2\sin(\omega t + \frac{\varphi_1 + \varphi_2}{2}) \cdot \cos(k_1 x + \frac{\varphi_1 - \varphi_2}{2}) . \end{cases} \quad (10)$$

In case of electromagnetic (transverse) waves there is analogous non-damping at $\omega > \omega_L$ solution [6], [7], [8], which being substituted into equations (1a), (3) leads to the traditional dispersion equation. The discussion on two branches of its solution (waves with over light phase velocity more $c$ and group velocity $d\omega/dk = c\sqrt{1 - \omega_L^2/\omega^2} < c$, and waves with velocity $\gtrsim \sqrt{v_x^2}$, also including collision damping) is present in papers [15], [16].

Dispersion relation for the high velocity branch has usual form

$$\frac{\omega}{k} \sim \frac{c}{\sqrt{1 - \omega_L^2/\omega^2}}, \quad (11)$$

where $c$ is light velocity. It ought to emphasize that for light wave, velocities are caused by the presence of the very small part of electrons with near light velocities, and such electrons would require special relativistic consideration.

In the case of low-speed branch dispersion relation has a trivially simple form [15]

$$\frac{\omega}{k} \gtrsim \sqrt{v_x^2}. \quad (12)$$

In the case of high-speed branch the critical point $v_x \simeq \omega/k$ corresponds to so high electron velocities $\sim c$ that due to exponential decreasing of electron distribution function at high velocities one can neglect these electrons. Therefore all set forth below in Section 5 about cutting off overtone amplitudes relates mainly to the low-velocity rarely mentioned branch, apparently not investigated enough experimentally. At that time for the high speed branch it might be necessary significant cutting off amplitudes $f_1$ at high velocities $v_z$ due to very low value of $f_0$.

The difficulties of turning $f_1$ into infinity $\pm \infty$ in the resonance point $v_x = \omega/k$ are kept also in Landau theory (where at small $|\text{Im } k|$ the function $|f_1|$ is finite but can be locally more $f_0$). They ought to be resolved naturally by the suitably carried out cutting off distribution function accounting for electron density finiteness and natural condition of the total distribution function ($f_0 + f_1$) to be positive in regions where kinetic equation becomes inapplicable due to growing $|f_1|$.

Accounting for periodical sign alternating of the perturbed part $f_1$ in $t$ or $x$ (now not small near the critical point) the natural request is bringing changes into kinetic equation in the singularity region of the approximate form

$$0 < f(v) \leq f_0 + |f_{1\max}| < 2f_0(v) \qquad (13)$$

(keeping in mind the possibility of including non-linear quadratic terms of wave equations what leads to appearance of overtones with $|f_1| \ll f_0$, see below). This request can be satisfied approximately with half-empirical successive cutting off values $f_1$ where condition (13) is not satisfied. As it will be shown below at the formal expansion of asymptotical solutions of the non-linearized wave equations in overtone series, violations of the condition (13) have place both for individual overtones and for their sum, which may require renormalization of $|f_{1\max}|$ (39).

This work is a further development of proposals which were formulated at the preliminary analysis of linearized equations [12]. The following sections are devoted to ways of solving non-linear wave equations. In Section 3 one considers longitudinal waves. Transverse waves are considered in Section 4. In Section 5 there are presented half-empirical proposals of ways of cutting off distribution functions. Results are resumed in Conclusion (Section 6).

The most precise behavior of the distribution function $f_1(v_x)$ near the singular point is determined by the inclusion of the collision terms in the kinetic equation. This eliminates the need for rules of Vlasov and Landau, as well as the existence itself of Landau damping as real physical phenomenon [12].

### 3. Expansion of solutions of non-linear equations of longitudinal waves in overtones

Assuming waves propagating in positive (to the right) direction to be

$$f_1 = \sum_{n=1} F_n \cos[n(\omega t - kx) + \varphi_n], \qquad (14)$$

$$E = \sum_{n=1} E_n \cos[n(\omega t - kx) + \psi_n], \qquad (15)$$

where $n = 1$ corresponds to the solutions (5)-(9) with an arbitrary value $E_1'$, substituting these ones into Eq. (1), (2) and equalizing coefficients at overtones with the same orders $n$, one obtains (at $e \equiv |e|$)

$$F_1(v, v_x) = \frac{eE_1}{m_e} \frac{\partial f_0/\partial v_x}{\omega - kv_x}, \qquad (16)$$

$$F_2(v, v_x) = \frac{eE_2}{2m_e} \frac{\partial f_0/\partial v_x}{\omega - kv_x} + \frac{e}{4m_e} \frac{E_1}{\omega - kv_x} \frac{\partial F_1}{\partial v_x} = \frac{e}{2m_e} E_2 \frac{\partial f_0/\partial v_x}{\omega - kv_x} + \frac{1}{4}\left(\frac{eE_1}{m_e}\right)^2 \frac{1}{\omega - kv_x} \frac{\partial}{\partial v_x} \frac{\partial f_0/\partial v_x}{\omega - kv_x}. \qquad (17)$$

Since

$$E_2 = \frac{-2\pi e n_e}{k} \int F_2 d\vec{v}, \qquad (18a)$$

then

$$E_2 = \frac{\pi e^2 n_e E_1}{2 m_e k} \frac{-\int \frac{\partial F_1/\partial v_x}{\omega - kv_x} d\vec{v}}{1 + \frac{\pi e^2 n_e}{k m_e} \int \frac{\partial f_0/\partial v_x}{\omega - kv_x} d\vec{v}} \qquad (18b)$$

at
$$\varphi_2 = 2\varphi_1 = \psi_2 - \pi/2 \qquad (18c)$$

and the dispersion equation in approximation of the 1th order

$$E_1 = \frac{-4\pi e n_e}{k} \int F_1 d\vec{v}. \qquad (19)$$

Analogous transforms for the 3-th order overtones lead to kinetic equation in the form

$$3(\omega - kv_x) F_3 \sin[3(\omega t - kx) + 3\varphi_1] = \frac{e}{m_e} E_3 \frac{\partial f_0}{\partial v_x} \sin[3(\omega t - kx) + 3\varphi_1] +$$

$$+ \frac{e}{2m_e} \frac{\partial}{\partial v_x} (E_1 F_2 + F_1 E_2) \sin[3(\omega t - kx) + 3\varphi_1], \qquad (20)$$

$$F_3(v, v_x) = \frac{e}{3m_e} E_3 \frac{\partial f_0/\partial v_x}{\omega - kv_x} + \frac{e}{6m_e} \frac{1}{\omega - kv_x} \frac{\partial}{\partial v_x} (E_1 F_2 + F_1 E_2), \qquad (21)$$

$$E_3 = \frac{-4\pi e n_e}{3k} \int F_3 d\vec{v}, \qquad (22)$$

which implies

$$E_3 = \frac{-\frac{2\pi e^2 n_e}{9 m_e k} \left( E_1 \int \frac{\partial F_2/\partial v_x}{\omega - kv_x} d\vec{v} + E_2 \int \frac{\partial F_1/\partial v_x}{\omega - kv_x} d\vec{v} \right)}{1 + \frac{4\pi e^2 n_e}{9 m_e k} \int \frac{\partial f_0/\partial v_x}{\omega - kv_x} d\vec{v}} \qquad (23)$$

with

$$\varphi_3 = 3\varphi_1 = \psi_3 - \pi/2. \qquad (24)$$

It follows from general physical considerations that a small physical convergence parameter of the overtone expansion appears parameter

$$\eta \lesssim \left| \frac{e}{m_e} \frac{E_1}{\sqrt{v_x^2}} \frac{\pi}{\omega} \right| \ll 1, \qquad (25)$$

that is the reached speed at acceleration of electrons by the field $E_1$ is much less the r.m.s electron velocity.

Expansion in overtones leads to expansion in decreasing amplitudes proportional $\eta^n$. Restricting with the first expansion terms one can details convergence condition (25). Thus, integrating in parts in continued integral intervals (which are in addition implied in the principal value sense) one obtains

$$E_2 \simeq -E_1 \left( \frac{eE_1}{m_e} \right) k \frac{4\omega^2 - \omega_L^2}{12\omega_L^4}, \qquad (26)$$

$$\eta \to \left| \frac{E_2}{E_1} \right| \lesssim \left| \left( \frac{eE_1}{m_e} \right) \frac{4\omega^2 - \omega_L^2}{12\omega_L^4} \frac{\omega}{\sqrt{v_x^2}} \right|, \qquad (27)$$

For all this it is supposed that the functions $F_n$ undergo a break with finite values equal to $\pm\sigma_n f_0$, $\sigma_n = \text{const}$ at different sides near the point $v_x = \omega/k$ according to described in Section 5 cutting off procedure. At calculating $E_n$ one neglected changing $F_n$ at cutting off procedures, although this can be account for at following calculations. The calculated values $E_n$ are the more precise, the narrower cut off intervals, that is smaller $E_1$.

All of the above can be similarly generalized to the transverse plasma waves in collisionless plasma. In this case dispersion equation has two branches of solutions: simultaneously fast electromagnetic wave and a relatively slow electron wave, respectively to $(\omega - kc) \sim 0$ or $(\omega - kv) \sim 0$ in the denominator of dispersion equation at velocity, respectively, $v \sim c$ or $v \sim \sqrt{v_x^2}$ where $\sqrt{v_x^2}$ is the rms velocity of electron distribution function.

### 4. Expansion of solutions of non-linear equations of transverse waves in overtones

In the case of transverse field $E = E_z(x,t)$, after substituting the expansions (14), (15) into the equations of transverse waves (1a), (3) and successively equalizing to zero sums of the coefficients at amplitudes of n-order overtones analogously to the foregoing procedure for longitudinal waves, one obtains (at $e \equiv |e|$)

$$F_1(v, v_x, v_z) = \frac{eE_1}{m_e} \frac{\partial f_0/\partial v_z}{\omega - kv_x}, \tag{28}$$

$$F_2(v, v_x, v_z) = \frac{eE_2}{2m_e} \frac{\partial f_0/\partial v_z}{\omega - kv_x} + \frac{eE_1}{4m_e} \frac{\partial F_1/\partial v_z}{\omega - kv_x} = \frac{eE_2}{2m_e} \frac{\partial f_0/\partial v_z}{\omega - kv_x} + \frac{1}{4}\left(\frac{eE_1}{m_e}\right)^2 \frac{\partial^2 f_0/\partial v_z^2}{(\omega - kv_x)^2}, \tag{29}$$

$$\begin{cases} 3(\omega - kv_x) F_3 \sin[3(\omega t - kx) + \varphi_3] = \frac{eE_3}{m_e} \frac{\partial f_0}{\partial v_z} \sin[3(\omega t - kx) + \varphi_3] + \\ + \frac{e}{2m_e}\left(E_1 \frac{\partial F_2}{\partial v_z} + E_2 \frac{\partial F_1}{\partial v_z}\right) \sin[3(\omega t - kx + \varphi_1)] + \frac{e}{2m_e}\left(E_2 \frac{\partial F_1}{\partial v_z} - E_1 \frac{\partial F_2}{\partial v_z}\right) \sin(\omega t - kx + \varphi_1). \end{cases} \tag{30}$$

The latter term gives small correcting addition $\sim E^3$ to the term with $n = 1$ and may be accounted for, followed by applying an recurrent iterative procedure, so after omitting it here, one ought to relate it to the first order term $n = 1$ with $\sin(\omega t - kx + \varphi_1)$, then one obtains finally

$$F_3(v, v_x, v_z) = \frac{eE_3}{3m_e} \frac{\partial f_0/\partial v_z}{\omega - kv_x} + \frac{e}{6m_e} \frac{E_1 \partial F_2/\partial v_z + E_2 \partial F_1/\partial v_z}{\omega - kv_x}, \tag{31}$$

with

$$\varphi_1 = \psi_1 - \pi/2, \quad \varphi_2 = 2\varphi_1 = \psi_2 - \pi/2, \quad \varphi_3 = 3\varphi_1 = \psi_3 - \pi/2 \tag{32a}$$

and dispersion equation in approximation of the 1th order

$$E_1 = \frac{-4\pi e n_e \omega}{\omega^2 - k^2 c^2} \int v_z F_1 d\vec{v}. \tag{32b}$$

The value $E_1$ is arbitrary. For $E_2$, $E_3$ one obtains

$$E_2 = \frac{2\pi e n_e \omega}{k^2 c^2 - \omega^2} \int v_z F_2 d\vec{v} = \frac{\dfrac{\pi e^2 n_e}{2m_e} \dfrac{\omega}{k^2 c^2 - \omega^2} E_1 \int v_z \dfrac{\partial F_1/\partial v_z}{\omega - kv_x} d\vec{v}}{1 - \dfrac{\pi e^2 n_e}{m_e} \dfrac{\omega}{k^2 c^2 - \omega^2} \int v_z \dfrac{\partial f_0/\partial v_z}{\omega - kv_x} d\vec{v}}, \qquad (33)$$

$$E_3 = \frac{4\pi e n_e \omega}{3(k^2 c^2 - \omega^2)} \int v_z F_3 d\vec{v} = \frac{2\pi e^2 n_e \omega}{9(k^2 c^2 - \omega^2)} \frac{\int v_z \dfrac{E_1 \partial E_2/\partial v_z + E_2 \partial F_1/\partial v_z}{\omega - kv_x} d\vec{v}}{1 - \dfrac{4\pi e^2 n_e \omega}{9 m_e (k^2 c^2 - \omega^2)} \int v_z \dfrac{\partial f_0/\partial v_z}{\omega - kv_x} d\vec{v}}. \qquad (34)$$

Due to antisymmetric in $v_z$ integrand function in the numerator of (35) with Maxwellian function $f_0$, it appears to be

$$E_2 = 0, \qquad (35)$$

thus the expression for $E_3$ will be

$$E_3 = \frac{\pi e^2 n_e \omega}{9 m_e (k^2 c^2 - \omega^2)} \left(\frac{eE_1}{m_e}\right)^2 E_1 \frac{\int v_z \dfrac{\partial^3 f_0/\partial v_z^3}{(\omega - kv_x)^3} d\vec{v}}{1 - \dfrac{4\pi e^2 n_e \omega}{9 m_e (k^2 c^2 - \omega^2)} \int v_z \dfrac{\partial f_0/\partial v_z}{\omega - kv_x} d\vec{v}} \qquad (36)$$

where integrals in $dv_x$ (excluding expression in the nominator of (33)) are implied in the principal value sense and can be evaluated using approximations of the type (13).

Note however that $F_2$ has no bound between positive and negative values at velocity $v_x = \omega/k$, and the principal value of the integral in $dv_x$ in the numerator of (33) equals $\infty$. Reasonable values of $F_2$ and $E_2$ with $E_2 \neq 0$ and a constrain on the distribution function $F_2$ can be found in natural manner by the procedure of cutting off $F_n$ based on condition (13) as it will be described in the following section.

Nonlinear term in the r.h.s. of Eq. (1a) with the two branches of solution of the linearized dispersion equation as a sum of both types of electromagnetic waves with amplitudes $F_1^{(1)}$ and $F_1^{(2)}$ leads in general case inevitably to appearance of hybrid terms of both branches. So, accounting for r.h.s. term in Eq. (1a), overtones expansion must include also hybrid terms with the coefficients determined after iterative substitution of this expansion into equations (1a) and (3).

It can be assumed that the general solution is the sum of two solution branches $f_1$ with the same boundary condition (at $x = 0$) of $E_1 \cos \omega t$ and $f_0$ for each branch.

### 5. Cutting off amplitudes of distribution functions[3]

Since in the kinetic equation we neglected term of collisions between all electrons, there is some uncertainty in the choice of ways to cutoff infinities of the function $|f_1(\vec{v})| \leq |f_{1\max}| < f_0(\vec{v})$. One of the ways is to represent a reasonable cutoff $|f_{1\max}|$ with the choice of interval $\pm \Delta v_x$ the left and right of the singularity point $(\omega - kv_x) = 0$, in which $f_1$ changes smoothly from $-|f_{1\max}|$ to $|f_{1\max}|$ (or vice versa) with the zero crossing in the middle part of the interval $-|\Delta v_x| \leftrightarrow |\Delta v_x|$ [13]. The same is keeping for the overtones (all of which are now finite), whereby there is a sequence of decreasing values

---

[3]On the use of constraints $|f_1| < f_0$ and values $f_1(v_x)$ near the point of singularity see also [13].

$F_{1\max}^{(n)}$ with necessity of renormalization left and right near the singularity point $|f_{1\max}(v_x)|$

$$|f_1| \to \alpha |f_1|, \quad \alpha = \left|F_{1\max}^{(1)}\right| \bigg/ \left|\sum_n F_{1\max}^{(n)}\right|. \tag{37}$$

It is expected that the choice of cutoff $F_{1\max}^{(n)}$ method effects very little on the dispersion relation.

The more convenient and general form of overtone expansion with more $n$ at Maxwellian $f_0$ can be presented (see [16]) as

$$E = \frac{1}{2} \{ \sum_{n>0} e^{-\delta x} \left[ E_n^{(s)} e^{in(\omega t - k_0 x)} \right] + \text{(complex conjugated)} \}; \tag{38}$$

$$F = \frac{1}{2} \{ \sum_{n>0} e^{-\delta x} \left[ F_n^{(s)} e^{in(\omega t - k_0 x)} \right] + \text{(complex conjugated)} \}, \tag{39}$$

in which $k_0$ is the real part of wave number; $\delta > 0$ is decrement, and $s$ is iteration order of collision damping.

### 6. Conclusion

Correspondingly to the general trend and principles exposed in [12], [13], [14] we have used an expansion of solutions of non-linearized equations as for longitudinal as well as transverse waves in ion-electron plasma in series of spontaneously arising non-damping overtones of the distribution function perturbation in the form

$$F = \sum F_n(\vec{v}) \cos \left[ n(\omega t - kx) + \varphi_n \right] \tag{40}$$

and the field

$$E = \sum E_n \cos \left[ n(\omega t - kx) + \psi_n \right], \tag{41}$$

$$\varphi_n = n\varphi_1, \quad \psi_n = \varphi_n + \pi/2. \tag{42}$$

There is verified the possibility of such expansion on principle with recurrent relations for expressing high order terms by the terms of lower orders. There is found striking phenomenon of naturally arising overtones at applying the only boundary harmonic electrical field $E_1$. Amplitudes $F_n(\vec{v})$ and $E_n$ decrease with increasing $n$ as powers of divergence parameter $\eta$ (25) or enough small boundary field $E_1$. There are obtained expressions for amplitudes of overtones $F_n$ and $E_n$ up to $n \leq 3$.

For solutions of linearized as well as non-linearized equations the required by definition reasonable condition for positivity $f_0 + f_1$ and anti-symmetry $\pm |f_1(v_x)|$ near the singularity points $v_x = \pm \omega/k$ of the total distribution function $0 \leq (f_0 + f_1) \leq 2f_0$ is violated both for longitudinal and transverse waves.

In accordance with the works [9], [14], [15] there arises a problem to introduce corrections into kinetic equation in the form of cutting off functions $\left| \sum F_n \right| \leq f_0$ for to satisfy the condition of positivity and amplitude constraints of the oscillatory distribution function. There are proposed preliminarily some half-empirical procedures of cutting off amplitudes of longitudinal and transverse waves, however they require proving and specifying them.

The oscillatory distribution functions $F_n(v_x)$ have jump from positive to negative values or vice versa near critical point $v_x = \pm \omega/k$.

The presented in some extent exotic results appear to be accessible to experimental examination as relatively to overtone features and existence of the low-velocity branch of transverse waves, as well as relatively to some unusual form of distribution function with a wave front at $v_x = \pm \omega/k$. However one ought to account for that really observable wave form is determined with addition of overtone waves with near frequencies in the wave packet, that is the resulting wave ought to be of some composed smoothed form traveling with the group velocity.

There is expected that detecting of overtones does not provide so much careful (as it had been noted in [12], [13]) measurements as, for example, that ones of Landau damping. Examples of the wave solutions for distribution functions in half-infinite slab problems with constrained integrals and the two additional boundary conditions: absence both fast backward waves and of kinematical waves, are presented in papers [18], [19]. See also [14], *Appendix 6*.